%% file: main.tex
\documentclass[runningheads]{llncs}
\usepackage[T2A]{fontenc}
\usepackage{array}
\usepackage{amsmath}

\usepackage{cite}
\usepackage{amsmath,amssymb,amsfonts}
\usepackage{algorithmic}
\usepackage{graphicx}
\usepackage{hyperref}
\usepackage{textcomp}
\usepackage{xcolor}
\usepackage{float}
\usepackage{caption}
\usepackage{subcaption}
\usepackage{lineno}
\usepackage{booktabs}
\usepackage{enumitem}
\usepackage{multirow}
\usepackage{hyperref}
\usepackage{graphicx}
\usepackage{xcolor}

\begin{document}
\title{Zero-Shot Domain Adaptation in CT Segmentation by Filtered Back Projection Augmentation}
\titlerunning{Zero-Shot Domain Adaptation in CT by FBP Augmentation}

\author{
    Talgat Saparov \inst{1, 3} \and
    Anvar Kurmukov \inst{2, 4} \and
    Boris Shirokikh  \inst{1} \and
    Mikhail Belyaev \inst{1}
}

% \author{
%     Talgat Saparov
%     Anvar Kurmukov
%     Boris Shirokikh \inst{1} \and
%     Ivan Zakazov \inst{1, 2} \and
%     Mikhail Belyaev \inst{1}
% }

%index{Shirokikh, Boris}
%index{Zakazov, Ivan}
%index{Belyaev, Mikhail}

\authorrunning{Anonymous Author et al}
% \authorrunning{B. Shirokikh, I. Zakazov et al}

\institute{
    Skolkovo Institute of Science and Technology, Moscow, Russia
    \and
    Artificial Intelligence Research Institute, Moscow, Russia
    \and
    Moscow Institute of Physics and Technology, Moscow, Russia
    \and
    Higher School of Economics, Moscow, Russia
    \\
    \email{saparov2130@gmail.com}
}
% \institute{
%     Skolkovo Institute of Science and Technology, Moscow, Russia
%     \and
%     Philips Research, Moscow, Russia
%     \\
%     \email{boris.shirokikh@skoltech.ru}
% }

\maketitle              % typeset the header of the contribution
\begin{abstract}
Domain shift is one of the most salient challenges in medical computer vision. Due to immense variability in scanners’ parameters and imaging protocols, even images obtained from the same person and the same scanner could differ significantly. We address variability in computed tomography (CT) images caused by different convolution kernels used in the reconstruction process, the critical domain shift factor in CT. The choice of a convolution kernel affects pixels’ granularity, image smoothness, and noise level. We analyze a dataset of paired CT images, where smooth and sharp images were reconstructed from the same sinograms with different kernels, thus providing identical anatomy but different style. Though identical predictions are desired, we show that the consistency, measured as the average Dice between predictions on pairs, is just $0.54$. We propose Filtered Back-Projection Augmentation (FPBAug), a simple and surprisingly efficient approach to augment CT images in sinogram space emulating reconstruction with different kernels. We apply the proposed method in a zero-shot domain adaptation setup and show that the consistency boosts from $0.54$ to $0.92$ outperforming other augmentation approaches. Neither specific preparation of source domain data nor target domain data is required, so our publicly released FBPAug\footnote{\url{https://github.com/STNLd2/FBPAug}} can be used as a plug-and-play module for zero-shot domain adaptation in any CT-based task.

\end{abstract}
\renewcommand{\arraystretch}{1.4}
\input{chapters/1_intro}
\input{chapters/2_method}

\input{chapters/3_experiments}

\input{chapters/4_results}

\input{chapters/5_conclusion}

\bibliography{bibliography}
\end{document}

%% file: chapters/1_intro.tex
\section{Introduction}

Computed tomography (CT) is a widely used method for medical imaging. CT images are reconstructed from the raw acquisition data, represented in the form of a sinogram. Sinograms are two-dimensional profiles of tissue attenuation as a function of the scanner's gantry angle. One of the most common reconstruction algorithms is Filtered Back Projection (FBP) \cite{schofield2020image}. This algorithm has an important free parameter called \textit{convolution kernel}. The choice of a convolution kernel defines a trade-off between image smoothness and noise level \cite{schaller2003spatial}. Reconstruction with a high-resolution kernel yields \textit{sharp} pixels and a high noise level. In contrast, usage of a lower-resolution kernel results in \textit{smooth} pixels and a low noise level. Depending on the clinical purpose, radiologists use different kernels for image reconstruction. 

Modern deep neural networks (DNN) are successfully used to automate computing clinically relevant anatomical characteristics and assist with disease diagnosis. However, DNNs are sensitive to changes in data distribution which are known as \textit{domain shift}.  Domain shift typically harms models' performance even for simple medical images such as chest X-rays \cite{zech2018variable}. In CT images, factors contributing to domain shift include \cite{kloenne2020domain}
slice thickness and inter-slice interval, different radiation dose, and reconstruction algorithms, e.g., FBP parameters. The latter problem is a subject of our interest.

Recently, several studies have reported a drop in the performance of convolutional neural networks (CNN), trained on \textit{sharp} images while being tested on \textit{smooth} images \cite{missert2019simulation,choe2019deep,lee2019ct}. Authors of \cite{sandfort2019data} proposed using generative adversarial networks (GAN) to generate realistic CT images imitating arbitrary convolution kernels. A more straightforward approach simultaneously proposed in \cite{missert2019simulation}, \cite{choe2019deep}, and \cite{lee2019ct} suggests using a CNN to convert images reconstructed with one kernel to images reconstructed with another.  Later, such image-to-image networks can be used either as an augmentation during training or as a preprocessing step during inference. 

%Such approaches are very intuitive: convolution operations used in CNN are the same as in FBP. However, these seemingly similar convolutions process different images: FBP operates on sinograms, whereas CNN layers work with reconstructed CT images. Moreover, local convolutions in sinogram space become global in the image space during reconstruction. In other words, to emulate a convolution kernel applied to sinograms, one needs to use a full CNN with a wide enough receptive field instead of a single convolutional layer. 

%Alternatively, several papers assess the impact of different, more straightforward augmentation techniques, including windowing \cite{kloenne2020domain,lee2018practical}, gamma correction \cite{tureckova2020improving}, image normalization \cite{gallardo2016normalizing}, and image filtering \cite{ohkubo2011image}. Two latter methods were specifically proposed to address the differences in the convolutional kernels using images in the pixel domain. Although these methods can be applied to any domain adaptation problem in a zero-shot setup, their simplicity limits the possible generalization effect. 

% 

%In this study, we aim to take the best of both approaches, achieving a high level of generalization with a physics-driven augmentation procedure. 
We propose FBPAug, a new augmentation method based on the FBP reconstruction algorithm. This augmentation mimics processing steps used in proprietary manufacturer's reconstruction software. We initially apply Radon transformation to all training CT images to obtain their sinograms. Then we reconstruct images using FBP but with different randomly selected convolution kernels. To show the effectiveness of our method, we compare segmentation masks obtained on a set of paired images, reconstructed from the same sinograms but with different convolutional kernels. These paired images are perfectly aligned; the only difference is their style: smooth or sharp. We make our code and results publicly available, so the augmentation could be easily embedded into any CT-based CNN training pipeline to increase its generalizability to smooth-sharp domain shift.

% The rest of the paper is structured as follows. In the second section we describe a FBPAug, a new augmentation method based on the FBP reconstruction algorithm. We also provide details on the datasets used in experiments and comparison augmentation approaches.
%\textbf{Our contribution} is two-fold. \textit{Firstly}, we propose FBPAug, a new augmentation method based on the FBP reconstruction algorithm. We initially apply Radon transformation to all training CT images to obtain their sinograms. Then we reconstruct images using FBP but with different randomly selected convolution kernels. This augmentation mimics processing steps used in proprietary manufacturer's reconstruction software. \reconsider{\textit{Secondly}, we decompose different domain shift aspects by analyzing a publicly available set of paired CT images reconstructed from the same sinogram with different convolutional kernels. These paired images are perfectly aligned; the only difference is their style: smooth or sharp. We train U-net using smooth kernels in a zero-shot setup and then show that the other augmentation approaches may result in inconsistent predictions, whereas FBPAug improves segmentation on the target domain and achieves high correspondence on smooth-sharp pairs while preserving the quality on the source domain. We make our code and results publicly available, so the augmentation could be easily embedded into any CT-based CNN training pipeline to increase its generalizability to smooth-sharp domain shift.}

%% file: chapters/2_method.tex
\section{Materials and methods}

In this section, we detail our augmentation method, describe quality metrics, and describe datasets which we use in our experiments.

\subsection{Filtered Back-Projection Augmentation}

Firstly, we give a background on a discrete version of inverse Radon Transform -- Filtered Back-Projections algorithm. FBP consists of two sequential operations: generation of filtered projections and image reconstruction by the Back-Projection (BP) operator.

Projections of attenuation map have to be filtered before using them as an input of the Back-Projection operator. The ideal filter in a continuous noiseless case is the ramp filter. Fourier transform of the ramp filter $\kappa(t)$ is $\mathcal{F}[\kappa(t)](w) = |w|$.
 
The image $I(x, y)$ can be derived as follows:
  \begin{equation}
  \label{eqn:main_eq}
      I(x, y) = \text{FBP}(p_\theta(t)) = \text{BP}(p_\theta(t) * \kappa(t)),
  \end{equation}
 where $*$ is a convolution operator, $t = t(x,y) = x\cos\theta + y\sin\theta$ and $\kappa(t)$ is the aforementioned ramp filter.
 
Assume that a set of filtered-projections $p_{\theta}(t)$ available at angles $\theta_1, \theta_2, ..., \theta_n$, such that $\theta_i = \theta_{i - 1} + \Delta\theta,~i=\overline{2,n}$ and $\Delta\theta = \pi / n$. In that case, BP operator transforms a function $f_\theta(t) = f(x\cos\theta + y\sin\theta)$ as follows:
  \[
  BP(f_\theta(t))(x, y) = \frac{\Delta\theta}{2\pi}\sum\limits_{i=1}^n f_{\theta_i}(x\cos\theta_i + y\sin\theta_i) = \frac{1}{2n}\sum\limits_{i=1}^n f_{\theta_i}(x\cos\theta_i + y\sin\theta_i)
  \]
 
In fact, $\kappa(t)$ that appears in (\ref{eqn:main_eq}) is a generalized function and cannot be expressed as an ordinary function because the integral of $|w|$ in inverse Fourier transform does not converge. However, we utilize the convolution theorem that states that $\mathcal{F}(f*g) = \mathcal{F}(f)\cdot\mathcal{F}(g)$. And after that we can use the fact that the BP operator is a finite weighted sum and Fourier transform is a linear operator as follows:
 \[
 \mathcal{F}^{-1}\mathcal{F}[I(x, y)] = \mathcal{F}^{-1}\mathcal{F}[\text{BP}(p_\theta * \kappa)] = \text{BP}(\mathcal{F}^{-1}\mathcal{F}[p_\theta * \kappa]) = \text{BP}(\mathcal{F}^{-1}\{\mathcal{F}[p_\theta]\cdot|w|\}),\]
 \[I(x, y) = \text{BP}\left(\mathcal{F}^{-1}\{\mathcal{F}[p_\theta]\cdot|w|\}\right).\]
  
However, in the real world, CT manufacturers use different filters that enhance or weaken the high or low frequencies of the signal. We propose a family of convolution filters $k_{a,b}$ that allows us to obtain a smooth-filtered image given a sharp-filtered image and vice versa. Fourier transform of the proposed filter is expressed as follows:
 \[\mathcal{F}[k_{a,b}](w) = \mathcal{F}[\kappa](w)(1 + a \mathcal{F}[\kappa](w)^b) = |w|(1 + a|w|^b).\]

Thus, given a CT image $I$ obtained from a set of projections using one kernel, we can simulate the usage of another kernel as follows:
\[\hat{I}(x, y) = \text{BP}\left(\mathcal{F}^{-1}\{\mathcal{F}[\mathcal{R}(I)]\cdot\mathcal{F}[k_{a,b}]\}\right).\]
 
\begin{figure}[b!]
    \centering
    \includegraphics[width=7cm]{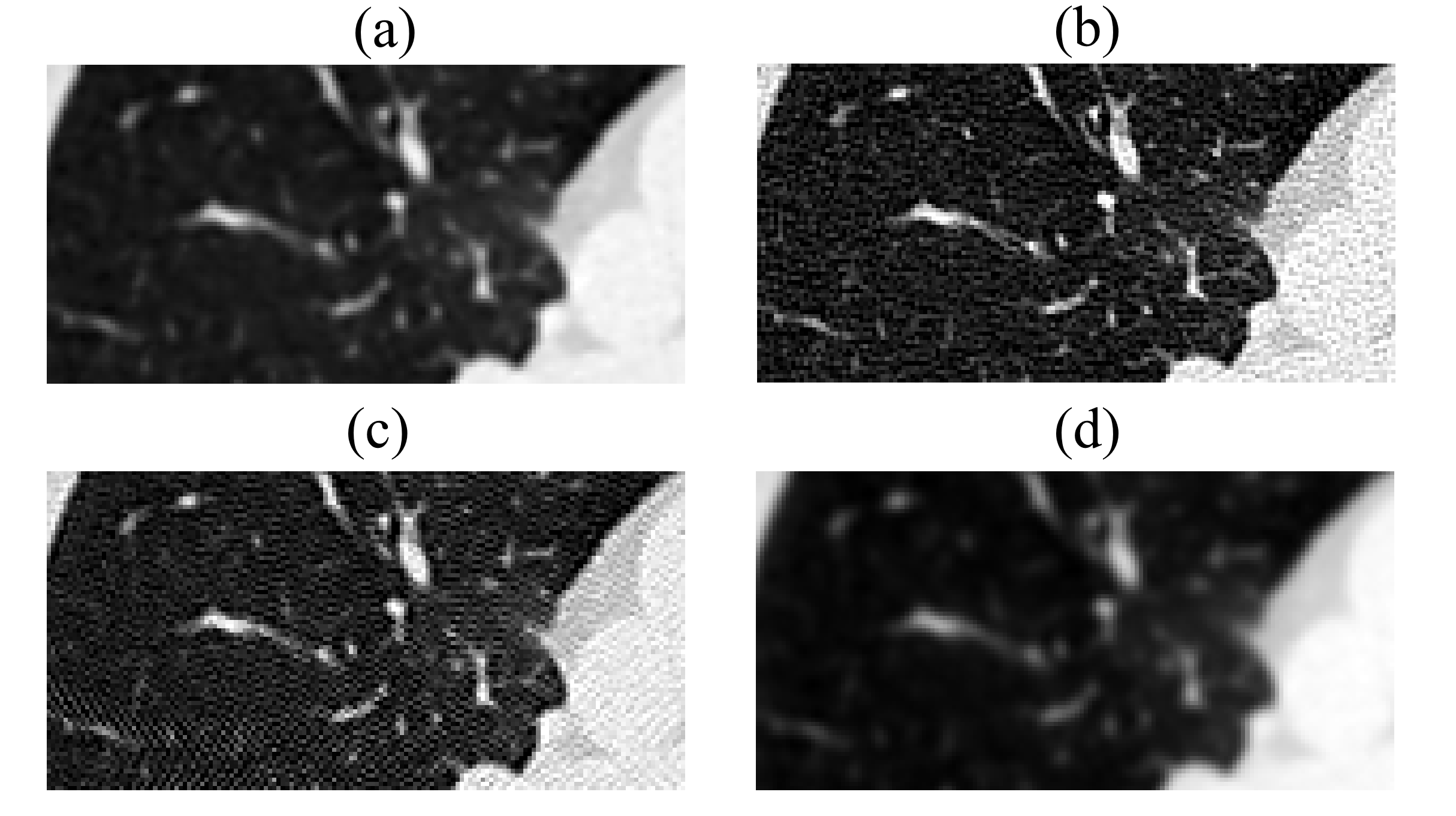}
    \caption{An example of paired CT slices (top row) and the effect of the augmentation by the proposed method (bottom row).  The top row contains original images: a slice reconstructed either with a smooth kernel (a) or sharp kernel (b). The bottom row shows augmented images: the top-left image processed by FBPAug with parameters $a=30,~b=3$ shifting it from smooth to sharp (c); the top-right image processed by FBPAug with parameters $a=-1,~b=0.7$ from sharp to smooth (d).}
    \label{fig:crops}
\end{figure}

Here, $a$ and $b$ are the parameters that influence the sharpness or smoothness of an output image and $\mathcal{R}(I)$ is a Radon transform of image $I$. The output of the Radon transform is a set of projections. Fig. \ref{fig:crops} shows an example of applying sharping augmentation on a soft kernel image (Fig. \ref{fig:crops}(a) to (c)) and vice versa: applying softening augmentation on a sharp kernel image (Fig. \ref{fig:crops}(b) to (d)).

\begin{figure}[!h]
\centering
\includegraphics[width=0.85\textwidth]{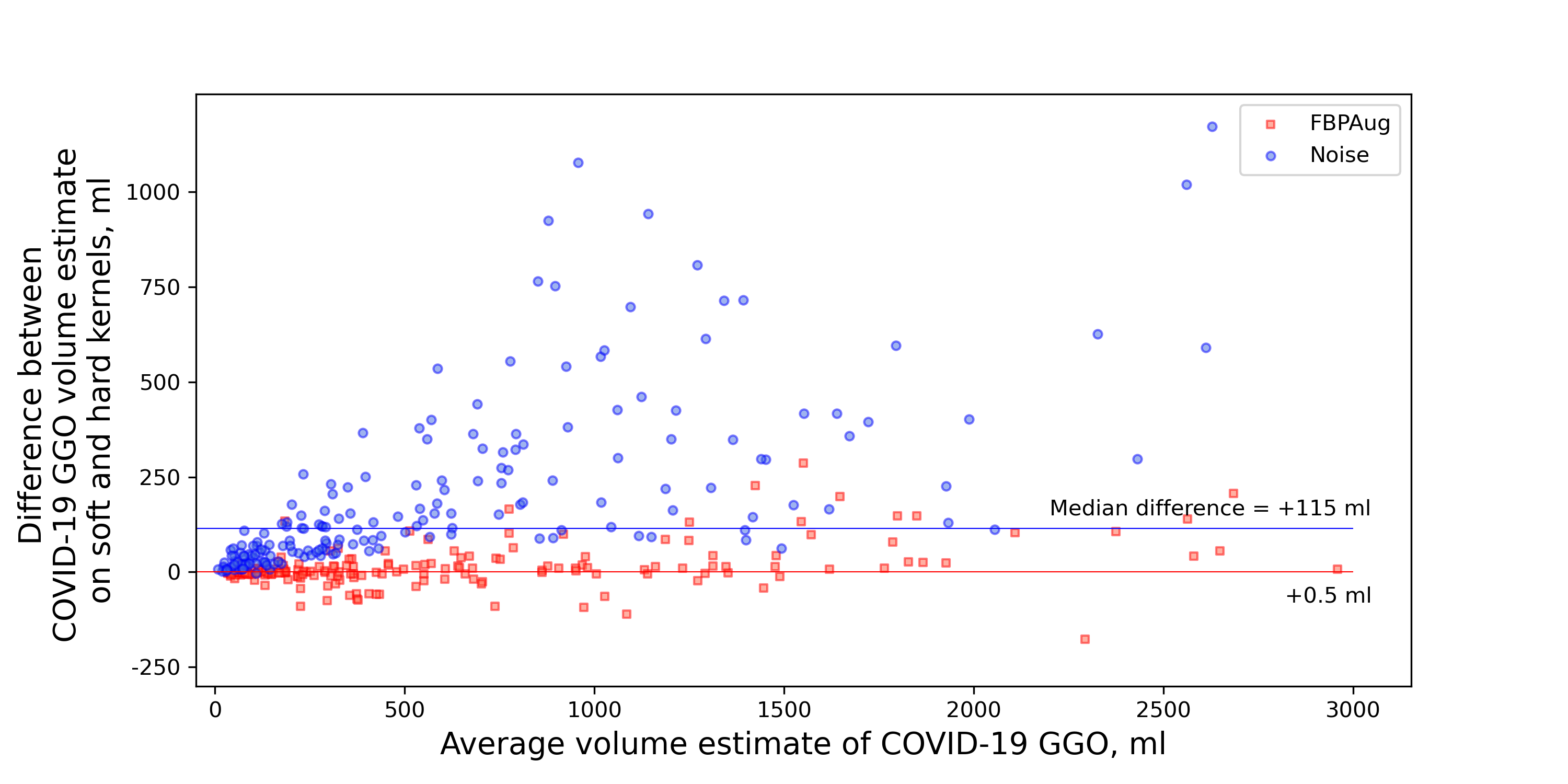}
\caption{Bland-Altman plot showing prediction agreement using FBPAug (proposed augmentation, red) and next best competitor (Gaussian noise, blue). Agreement is measured between predictions on paired images reconstructed with soft and hard convolution kernels from \textit{Covid-private} dataset. Difference in image pairs were always computed as $\text{Volume}_{\text{soft}} - \text{Volume}_{\text{sharp}}$.}
\label{fig:bland-altman}
\end{figure}
\subsection{Comparison augmentation approaches}

We compare the proposed method with three standard augmentations: gamma transformation (Gamma), additive Gaussian noise (Noise), and random windowing (Windowing), the technique proposed by \cite{kloenne2020domain}. As a baseline method, we train a network without any intensity augmentations (Baseline).

\vspace{0.2cm}
\noindent \textbf{Gamma} \cite{tureckova2020improving}, augments images using gamma transformation: 
\[\hat{I}(x, y) = \left(\frac{I(x, y) - m}{M - m}\right)^\gamma \cdot(M - m) + m,\]
where $M = \max(I(x, y))$, $m=\min(I(x, y))$ with a parameter $\gamma$, such that we randomly sample logarithm of $\gamma$ from $\mathcal{N}(0, 0.2)$ distrubution. 

\vspace{0.2cm}
\noindent\textbf{Noise} is the additive gaussian noise from $\mathcal{N}(0, 0.1)$ distribution. 

\vspace{0.2cm}
\noindent \textbf{Windowing} \cite{kloenne2020domain} make use of the fact that different tissue has diferent attenuation coefficient. We uniformly sample the center of the window $c$ from $[-700, -500]$ Hounsfield units (HU)  and the width of the window $w$ from $[1300, 1700]$ HU. Then we clip the image to the $[c - w/2, c + w/2]$ range.
% using following formula:
% %$\mathcal{U}([-700, -500])$, $\mathcal{U}([1300, 1700])$
% \begin{equation*}
% \hat{I}(x, y) = 
%  \begin{cases}
%   c - w/2, &\text{if } I(x, y) \le c - w/2 \\
%   I(x, y, z), &\text{if } I(x, y) \in (c - w/2, c + w/2) \\
%   c + w/2, &\text{if } I(x, y) \ge c + w/2
%  \end{cases}
% \end{equation*}

\noindent \textbf{FBPAug} parameters were sampled as follows. We uniformly sample $a$ from $[10.0, 40.0]$ and  $b$ from $[1.0, 4.0]$ in sharpening case and $a$ from $[-1.0, 0]$, $b$ from $[0.1, 1.0]$ in smoothing case. 

In all experiments, we zoom images to $1 \times 1$ mm pixel size and use additional rotations and flips augmentation. With probability $0.5$ we rotate an image by multiply of $90$ degrees and flip an image horizontally or vertically. 

\subsection{Datasets}

We report our results on two datasets: \textit{Mosmed-1110} and a private collection of CT images with COVID-19 cases (\textit{Covid-private}). Both datasets include chest CT series (3D CT images) of healthy subjects and subjects with the COVID-19 infection.

\paragraph{Mosmed-1110}
\label{par:data:1110} The dataset consists of 1110 CT scans from Moscow clinics collected from 1st of March, 2020 to 25th of April, 2020 \cite{morozov2020mosmeddata}. The original images have $0.8$ mm inter-slice distance, however the released studies contain every 10th slice so the effective inter-slice distance is $8$ mm. \textit{Mosmed-1110} contains only $50$ CT scans that are annotated with the binary masks of ground-glass opacity (GGO) and consolidation. We additionally ask three experienced radiologists to annotate another $46$ scans preserving the methodology of the original annotation process. Further, we use the total of $96$ annotated cases from \textit{Mosmed-1110} dataset. 

\paragraph{Covid-private}
\label{par:data:private}

All images from Covid-private dataset are stored in the DICOM format, thus providing information about corresponding convolution kernels. The dataset consists of paired CT studies ($189$ pairs in total) of patients with COVID-19. In contrast with many other datasets, all of studies contain two series (3D CT images); the overall number of series is $378$. Most importantly, every pair of series were obtained from one physical scanning with different reconstruction algorithms. It means the slices within these images are perfectly aligned, and the only difference is \textit{style} of the image caused by different convolutional kernels applied. 
%Data include paired images with the following kernels. GE manufacturer: SOFT, LUNG (40); LUNG, STANDARD (21).  Toshiba manufacturer: FC07, FC55 (40); FC07, FC51 (27).  Siemens manufacturer: B31f, B70f (38); B31s, B60s (23). First kernel in each pair is soft, second is sharp. 
\textit{Covid-private} does not contain ground truth mask of GGO or consolidation, thus we only use it to test predictions agreement.

% \todo{SAMPLE TEXT} 0.7 - 1.25 mm spacing private vs 8mm spacing mosmed-1110

\subsection{Quality metrics}

For the comparison, we use the standard segmentation metric, Dice Score. Dice Score (DSC) of two volumetric binary masks $X$ and $Y$ is computed as $DSC = \frac{2|X \cup Y|}{|X| + |Y|}$,
where $|X|$ is the cardinality of a set $X$.
% = \sum_{ijk} x_{ijk}

Furthermore, we perform statistical analysis ensure significance of the results.
We use one-sided Wilcoxon signed-rank test as we consider DSC scores for two methods are paired samples. 
To adjust for multiple comparisons we use Bonferroni correction.
%The null hypothesis of Wilcoxon test is $H_0: \mathbb{P} (X > Y) = \mathbb{P} (X < Y)$, and the alternative is $H_1: \mathbb{P} (X > Y) > \mathbb{P} (X < Y)$. Here, $\mathbb{P} (X > Y)$ is the probability of an observation from population $X$ exceeding an observation from population $Y$.

%To compare if different augmentations result in significantly better or worse results in comparison with ground truth we use  two one-sided T-test (TOST). We assume DSC scores for two methods to be independently generated samples, thus we can apply TOST to test the inequality of distributions. 
%The null hypothesis of TOST is $H_0: \mu_U - \mu_L < \Delta_L \text{ or } \mu_U - \mu_L > \Delta_U$. The alternative hypothesis is $H_1: \Delta_L < \mu_U - \mu_L < \Delta_U$. Here, $\Delta_U$ and $\Delta_L$ are the upper and lower equivalence bounds, respectively. We set these bounds to the $\pm 1/4$ of the estimated standard deviation.

%We compare methods' predictions with each other on the paired images. We consider DSC scores for two methods as paired samples and use one-sided Wilcoxon signed-rank test to to test if our method performs significantly better. 
%The null hypothesis of Wilcoxon test is $H_0: \mathbb{P} (X > Y) = \mathbb{P} (X < Y)$, and the alternative is $H_1: \mathbb{P} (X > Y) > \mathbb{P} (X < Y)$. Here, $\mathbb{P} (X > Y)$ is the probability of an observation from population $X$ exceeding an observation from population $Y$.

%% file: chapters/3_experiments.tex
\section{Experiments}

\subsection{Experimental pipeline}

To evaluate our method, we conduct two sets of experiments for COVID-19 segmentation.

First, we train five separate segmentation models: baseline with no augmentations, FBPAug, Gamma, Noise, and Windowing on a \textit{Mosmed} dataset to check if any augmentation results in significantly better performance. \textit{Mosmed} is stored in Nifti format and does not contain information about the kernels. Thus, we use it to estimate the in-domain accuracy for COVID-19 segmentation problem.

\begin{figure}[!h]
\centering
\includegraphics[width=9.5cm]{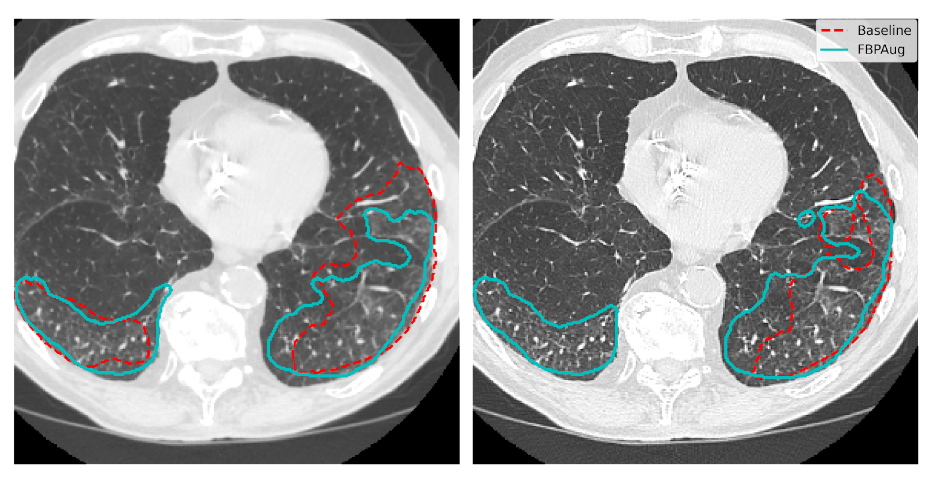}
\caption{Example prediction. Left - example image reconstructed with a \textit{smooth} kernel. Right - image  reconstructed with a \textit{sharp} kernel from the same sinogram. Red dashed contour - the baseline model (predictions agreement is $0.66$), blue contour - FBPAug (predictions agreement is $0.93$).}
\label{fig:slices-contours}
\end{figure}

Second, we use trained models from the previous experiment to make predictions on a paired \textit{Covid-private} dataset. We compare masks within each pair of sharp and soft images using Dice score to measure prediction agreement for the isolated domain shift reasons, as the only difference between images within each pair is their smooth or sharp style, see Fig. \ref{fig:slices-contours}.

\subsection{Network architecture and training setup}

For all our experiments, we use a  slightly modified 2D U-Net \cite{unet}. We prefer the 2D model to 3D since in the Mosmed-1110 dataset images have an 8 mm inter-slice distance and the inter-slice distance of Covid-private images is in the range from 0.8 mm to 1.25 mm. Furthermore, the 2D model shows performance almost equal to the performance of the 3D model for COVID-19 segmentation \cite{goncharov2020ct}. In all cases, we train the model for $100$ epochs with a learning rate of $10^{-3}$. Each epoch consists of $100$ iterations of the Adam algorithm \cite{kingma2014adam}. 
% bland-altman
% stat: to show our method achieves statistically indistinguishable results with the best of the other methods

\begin{table}[!h]
\caption{Comparison results. Numbers are mean (std) obtained on 3-fold cross-validation. Results for \textit{Mosmed-1110} are segmentation Dice score compared with ground truth; for \textit{Covid-private} are predictions agreement (between paired images) measured using Dice score.}
% The best results in each column are shown in bold.
\begin{center}
\begin{tabular}{| l | c | c | c | c | c |}
% \cmidrule(lr){2-5}
\hline
& Baseline & FBPAug & Gamma & Noise & Windowing \\
% \cmidrule(lr){2-3} \cmidrule(lr){4-5} \cmidrule(lr){6-7}
%   Experiment \ \  & Dice  \    & Surface dice 1.0       \    & Dice  \       & Surf. dice 1.0       \ & Dice    \    & Surf. dice 1.0 \\
\hline
Mosmed-1110          &$~0.56 ~(0.23)~$  &    $~0.59 ~(0.22)~$&    $\bf~0.61 ~(0.19)~$&    $~0.56 ~(0.21)~$&    $~0.59 ~(0.18)~$\\
Covid-private         &$0.54 ~(0.27)$  &    $\bf0.92 ~(0.05)$&    $0.68 ~(0.21)$&    $0.79 ~(0.13)$&    $0.63 ~(0.23)$\\

\hline
\end{tabular}
\end{center}
\label{tab2:cancer-results-gt}
\end{table}
At each iteration, we sample a batch of 2D images with batch size equals to $32$. %whole slices of
The training was conducted on a computer with 40GB NVIDIA Tesla A100 GPU. It takes approximately $5$ hours for the experiments to complete.

%% file: chapters/4_results.tex
\section{Results}
\label{sec:results}

Tab. \ref{tab2:cancer-results-gt} summarizes our results. First, experiments on \textit{Mosmed-1110} show that segmentation quality almost does not differ for compared methods. The three best augmentation approaches are not significantly different (p-value for Wilcoxon test are 0.17 for FBPAug vs Gamma and 0.71 for FBPAug vs Windowing). Thus our method does not harm segmentation performance. Our segmentation results are on-par with best reported for this dataset \cite{goncharov2020ct}. Next, we observe a significant disagreement in predictions on paired (\textit{smooth} and \textit{sharp}) images for all methods, except FBPAug (p-values for Wilcoxon test for FBPAug vs every other method are all less than $10^{-16}$). For FBPAug and its best competitor, we plot a Bland-Altman plot, comparing GGO volume estimates Fig. \ref{fig:bland-altman}. We can see that the predictions of FBPAug model agree independent of the volume of GGO.

%% file: chapters/5_conclusion.tex
\section{Conclusion}

We propose a new physics-driven augmentation methods to eliminate domain shifts related to the usage of different convolution kernels. It outperforms existing augmentation approaches in our experiments. We release the code, so our flexible and ready-to-use approach can be easily incorporated into any existing deep learning pipeline to ensure zero-shot domain adaptation.

The results have been obtained under the support of the Russian Foundation for Basic Research grant 18-29-26030.